\documentclass[aps,prb,reprint,a4paper,floatfix,groupedaddress,amsmath,amsfonts,amssymb]{revtex4-1}
\usepackage{mathptmx}
\newcommand{\upperRomannumeral}[1]{\uppercase\expandafter{\romannumeral#1}}
\usepackage[utf8]{inputenc} 
\usepackage{graphicx}
\usepackage{textcomp}
\usepackage{xspace}
\usepackage{url}
\usepackage[
bookmarks=true,         
    pdffitwindow=false,     
    colorlinks=true,       
    linkcolor=red,          
    citecolor=blue,        
    filecolor=magenta,      
    urlcolor=blue           
]{hyperref}
\usepackage{xcolor,soul}
\usepackage[
,textwidth=17.5cm
,textheight=23.5cm
,verbose
,dvips
]{geometry}

\DeclareMathAlphabet{\mathcal}{OMS}{cmsy}{m}{n}

\begin{document}
\title{Investigating the origin of the nonradiative decay of bound excitons in GaN nanowires}
\author{Christian Hauswald}
\email[Author to whom correspondence should be addressed. Electronic mail: ]{hauswald@pdi-berlin.de}
\author{Pierre Corfdir}
\author{Johannes K. Zettler}
\author{Vladimir M. Kaganer}
\author{Karl K. Sabelfeld}
\email[Permanent address: Institute of Computational Mathematics and Mathematical Geophysics, Russian Academy of Sciences, and Novosibirsk State University, 
Lavrentiev Prosp. 6, 630090 Novosibirsk, Russia]{}
\author{Sergio Fern\'{a}ndez-Garrido}
\author{Timur Flissikowski}
\author{Vincent Consonni}
\email[Present address: LMGP, Univ. Grenoble Alpes - CNRS, 38000 Grenoble, France]{}
\author{Tobias Gotschke}
\email[Present address: OSRAM Opto Semiconductors GmbH, Leibnizstr. 4, 93055 Regensburg, Germany]{}
\author{Holger T. Grahn}
\author{Lutz Geelhaar}
\author{Oliver Brandt}
\affiliation{Paul-Drude-Institut für Festkörperelektronik,
Hausvogteiplatz 5--7, 10117 Berlin, Germany}

\date{\today}
\begin{abstract}
We investigate the origin of the fast recombination dynamics of bound and free excitons in GaN nanowire ensembles by temperature-dependent  photoluminescence spectroscopy using both continuous-wave and pulsed excitation. The exciton recombination in the present GaN nanowires is dominated by a nonradiative channel between 10 and 300\,K. Furthermore, bound and free excitons in GaN NWs are strongly coupled even at low temperatures resulting in a common lifetime of these states. By solving the rate equations for a coupled two-level system, we show that one cannot, in practice, distinguish whether the nonradiative decay occurs directly via the bound or indirectly via the free state. The nanowire surface and coalescence-induced dislocations appear to be the most obvious candidates for nonradiative defects, and we thus compare the exciton decay times measured for a variety of GaN nanowire ensembles with different surface-to-volume ratio and coalescence degrees. The data are found to exhibit no correlation with either of these parameters, i.\,e., the dominating nonradiative channel in the GaN nanowires under investigation is neither related to the nanowire surface, nor to coalescence-induced defects for the present samples. Hence, we conclude that nonradiative point defects are the origin of the fast recombination dynamics of excitons in GaN nanowires.

\end{abstract}

\maketitle
\section{Introduction}
GaN nanowires (NWs) offer novel opportunities for the integration of optically active nanostructures based on GaN with electronic components based on silicon.\cite{Kikuchi2004, Sekiguchi2008c, Gonzalez-Posada2012a, Brubaker2013a} In contrast to planar GaN layers, they can be fabricated with a low density of structural defects directly on Si substrates by molecular-beam epitaxy (MBE).\cite{Sanchez-Garcia1998, Kikuchi2004, Geelhaar2011a} Because of the absence of threading dislocations, exciton recombination in GaN NWs is assumed to be predominantly radiative.\cite{Das2011} However, this expectation is inconsistent with the experimentally observed short decay times of bound excitons in GaN NWs. Even at low temperatures, time-resolved photoluminescence (TRPL) measurements yield typically~\cite{*[{With the sole exception of the work by }] [{, who determined decay times between 500\,ps and 1\,ns for GaN NWs with exceptionally large dimensions (up to $22\times 1.2$\,\textmu m$^2$) using very high excitation densities (up to 190\,\textmu J/cm$^2$).}] Schlager2011} decay times in the range of a few tens to about 200\,ps,\cite{Calleja2000, Corfdir2009d, Park2009, Park2010, Gorgis2012, Hauswald2013, Hauswald2014} i.\,e., significantly shorter than the radiative lifetime of the bound exciton state in bulk GaN of at least 1\,ns.\cite{Monemar2008,Monemar2010} Since NWs inevitably possess a large surface-to-volume ratio, it is often suspected that nonradiative surface recombination causes these fast decay times.\cite{Corfdir2009d, Park2009, Park2010, Korona2012, Gorgis2012} Recent experimental results, however, show that also comparatively thick NWs exhibit short decay times which cannot be ascribed to surface recombination alone.\cite{Hauswald2013, Hauswald2014}

Coalescence of NWs is a phenomenon less frequently considered as the possible origin of nonradiative recombination in GaN NWs. Given that the NW density in spontaneously formed GaN NW ensembles is on the order of $10^9$ to $10^{10}$\,cm$^{-2}$, coalescence of adjacent NWs during their formation is inevitable. Basal-plane stacking faults\cite{Consonni2009, Grossklaus2013a} and chains of dislocations\cite{Grossklaus2013a} have been observed as consequences of the coalescence of GaN NWs. The former defects capture free excitons efficiently, but manifest themselves by characteristic radiative transitions at specific energies.\cite{Chen2006b, Lefebvre2011, Nogues2014, Laehnemann2014} The latter, in contrast, were found to act as efficient nonradiative centers for excitons.\cite{Consonni2009} 

In the present paper, we investigate the radiative and nonradiative decay of free and bound excitons in GaN NW ensembles grown by molecular-beam epitaxy (MBE) on Si(111) substrates. We first focus on a representative sample to demonstrate that the exciton lifetime is dominated by a nonradiative decay channel between 10 and 300\,K. We also show that the donor-bound and free exciton states exhibit a common effective lifetime revealing an efficient coupling of both states even at low temperatures. An important consequence of this coupling is the fact that situations where the nonradiative decay occurs either via the bound or the free states are practically indistinguishable, since they result in the same decay time for the coupled system. Next, we compare the effective exciton lifetimes at low temperatures measured for 19 different GaN NW ensembles with different surface-to-volume ratio and coalescence degree. No systematic variation with either of these quantities is observed, suggesting that the dominating nonradiative channel in the present GaN NWs is neither related to the surface nor to coalescence-induced defects. 

\section{Experimental Details}
The investigated GaN NW ensembles were grown by plasma-assisted MBE on Si(111) substrates relying on either their spontaneous formation under suitable conditions\cite{Geelhaar2011a, Garrido2012} or employing selective-area growth (SAG) on a thin AlN buffer layer deposited on the Si substrate prior to NW growth.\cite{Schumann2011b} The substrate temperatures were in the range of 780 to 830\,°C, and the Ga/N flux ratio was changed accordingly to guarantee effectively N-rich conditions.\cite{Garrido2013}
 
The morphological properties of the as-grown GaN NW ensembles were studied by field-emission scanning electron microscopy as described in detail in Ref.~\onlinecite{Brandt2014}. In particular, top-view micrographs were used to determine the area and perimeter of several hundreds of individual NWs for each of the samples under investigation.

In order to investigate the optical properties of the GaN NWs, the as-grown samples were mounted onto the cold-finger of a microscope cryostat. Continuous-wave photoluminescence (cw PL) spectroscopy was performed by exciting the ensembles with the 325\,nm (3.814\,eV) line of a He-Cd laser focused onto the sample with an excitation density of less than 1\,W/cm$^2$, leading to a photogenerated carrier density below $10^{14}$\,cm$^{-3}$. The PL signal was spectrally dispersed by a 80\,cm monochromator providing a spectral resolution of 0.25\,meV and detected with a cooled charge-coupled device array. The TRPL measurements were performed by exciting the ensemble with the second harmonic (325\,nm) of pulses with a duration of about 200\,fs from an optical parametric oscillator synchronously pumped by a femtosecond Ti:sapphire laser, which itself was pumped by a frequency-doubled Nd:YVO$_4$ laser. A low excitation density corresponding to an energy fluence of 1\,\textmu J/cm$^2$ per pulse was chosen to prevent the saturation of impurities such as neutral donors or possible nonradiative point defects during the measurements. Assuming that all incident light is absorbed by the NWs, the upper limit of the photogenerated carrier density is $5\times 10^{16}$\,cm$^{-3}$. The transient PL signal was dispersed by a monochromator providing a spectral resolution of 4\,meV and detected by a streak camera with a temporal resolution of 50\,ps. In all cases, the excited area on the as-grown samples included at least 100 NWs.

\section{Results and discussion}
\subsection{Diameter and coalescence degree}
\label{sec:structure}

\begin{figure}[t]
   \includegraphics[width=7.9cm]{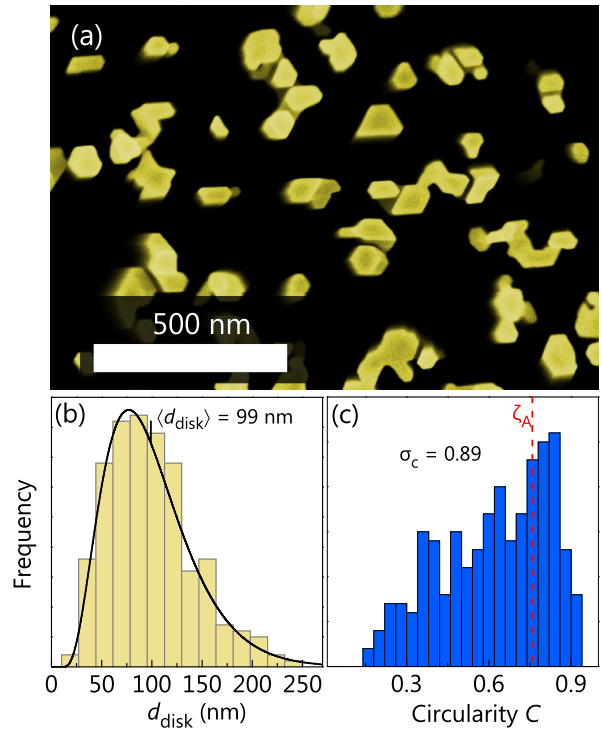}
   \caption{\label{fig:fig1}(a)(color online) Top-view scanning electron micrograph of the representative GaN NW ensemble (sample R). (b) Histogram representing the distribution of equivalent disk diameters $d_{\text{disk}}$ calculated from the NW top facet area $A$. A fit with a shifted Gamma distribution to the data yields $\langle d_{\text{disk}}\rangle=99$\,nm. (c) Histogram showing the circularity $C$ of 265 randomly selected NW top facets. Employing a threshold for the circularity of $\zeta_{\text{A}}=0.762$ (vertical dashed line) in order to distinguish between uncoalesced NWs and coalesced aggregates, a coalescence degree of $\sigma_{\text{c}} = 0.89$ is obtained.\cite{Brandt2014}}\end{figure}

Figure~\ref{fig:fig1}(a) depicts a top-view scanning electron micrograph of a representative, spontaneously formed GaN NW ensemble on Si (hereafter called 'sample R'). The GaN NWs have a length of $(2.3\pm 0.1)$\,\textmu m (not shown) and a density of 4.6$\times$10$^9$\,cm$^{-2}$. By measuring the area $A$ of their top facets, the distribution of equivalent disk diameters $d_{\text{disk}}=2 \sqrt{A/\pi}$ is extracted as depicted in the histogram in Fig.~\ref{fig:fig1}(b). The fit of these data with a shifted Gamma distribution yields a mean equivalent disk diameter $\langle d_{\text{disk}}\rangle=99$\,nm. 

The asymmetry of this distribution is a direct manifestation of the NW coalescence in the ensemble.\cite{Brandt2014} As seen in Fig.~\ref{fig:fig1}(a), the ensemble contains predominantly NWs with branched, kinked, and elongated top facets, i.\,e., shapes that differ strongly from the regular hexagonal characteristic for single NWs. These shapes originate from the coalescence of NWs formed in close vicinity.\cite{Brandt2014} As proposed in Ref.~\onlinecite{Brandt2014}, we employ the circularity $C$ of the plan-view shape of the NWs as a measure for the coalescence degree of the ensemble. For an arbitrary, contractible two-dimensional shape (i.\,e., one without any holes), $C$ is defined by $C=4\pi A/P^2$ with the area $A$ and the perimeter $P$. A circle has the maximum circularity of $C=1$, and a regular hexagon is characterized by $C=0.907$. NW coalescence results inevitably in shapes with significantly lower values for $C$.

Figure \ref{fig:fig1}(c) displays a histogram of the circularity of 265 NWs from sample R. The distribution is rather continuous with a maximum at $C=0.8$. Following Ref.~\onlinecite{Brandt2014}, we consider NWs with $C<\zeta_{\text{A}}=0.762$ as coalesced [cf.\ dashed line in Fig.~\ref{fig:fig1}(c)]. The coalescence degree $\sigma_{\text{C}}$ of the GaN NW ensemble is then given by:\cite{Brandt2014}
\begin{align}
\sigma_{\text{C}} = \frac{A_{C<\zeta_{\text{A}}}}{A_\text{T}} = 0.89.
\end{align}
Here, $A_{C<\zeta_{\text{A}}}$ represents the sum of all top-facet areas with $C<\zeta_{\text{A}}$, and $A_\text{T}$ denotes the total area of all NW top facets. Since the length of the GaN NWs (see above) is fairly homogeneous and much larger than their diameter, $\sigma_{\text{C}}$ directly represents the volume belonging to NWs which have participated in at least one coalescence event during growth.

\subsection{Low-temperature photoluminescence}
\label{sec:LTPL}
\begin{figure}[t]
   \includegraphics[width=8.2cm]{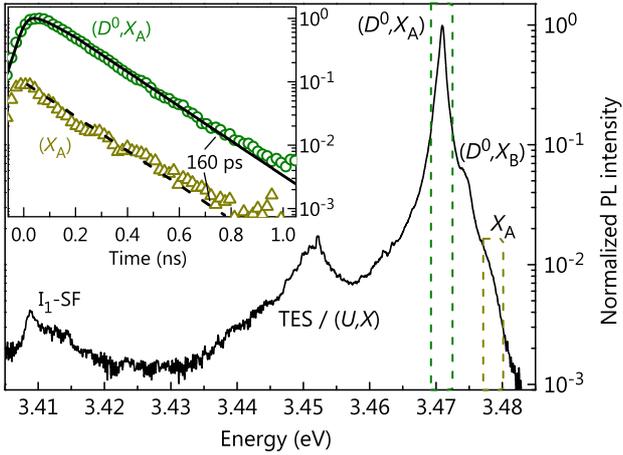}
  \caption{\label{fig:fig2} Low-temperature (10\,K) cw PL spectrum of sample R. The spectrum is dominated by the recombination of A excitons bound to neutral donors [($D^0,X_{\text{A}}$)] at 3.471\,eV. Transitions attributed to the recombination of B-excitons bound to neutral donors [($D^0,X_{\text{B}}$)] at 3.475\,eV and free A excitons ($X_{\text{A}}$) at 3.478\,eV are also detected. The origin of the TES/($U,X$) band at 3.45\,eV is discussed in the text. The dashed boxes indicate a 3\,meV wide spectral window centered around the ($D^0,X_{\text{A}}$) and $X_{\text{A}}$ transitions used to spectrally integrate the low-temperature TRPL transients depicted in the inset. The solid line in the inset represents a fit to the ($D^0,X_{\text{A}}$) transient using a single exponential decay convoluted with the system response function yielding an effective decay time of 0.16\,ns within the first ns. The dashed line is a guide to the eye and shows that the $X_{\text{A}}$ transient has the same effective lifetime as the ($D^0,X_{\text{A}}$) state within the error margin of the fit.}
\end{figure}

Figure~\ref{fig:fig2} shows the cw PL spectrum of sample R obtained at 10\,K. The spectrum is dominated by the recombination of A excitons bound to neutral donors [($D^0,X_{\text{A}}$)] at 3.471\,eV with a full width at half maximum of 1\,meV. On the high energy side of this transition, a shoulder is visible at 3.475\,eV, which stems from the recombination of B excitons bound to neutral donors [($D^0,X_{\text{B}}$)]. The transition arising from the radiative recombination of free A excitons ($X_{\text{A}}$) is centered at 3.478\,eV and is approximately two orders of magnitude less intense than the ($D^0,X_{\text{A}}$) line, reflecting the low excitation density used in these experiments. At 3.45\,eV, a set of closely spaced, narrow lines is observed. Besides the intrinsic two-electron satellite (TES) of the ($D^0,X_{\text{A}}$) transition in GaN,\cite{Dean1967, Monemar2008} several additional transitions have been reported in this energy range for GaN NWs,\cite{Calleja2000} which are related to excitons bound to as yet unknown defects [($U$,X)]  possibly related to the NW surface.\cite{Brandt_prb_2010, Lefebvre2011} Additionally, acceptor-bound excitons have been reported in this energy range.\cite{Monemar_jpcm_2001, Monemar2009} Finally, several lines originating from the recombination of excitons bound to $I_1$ basal-plane stacking faults (SF) are detected at 3.41\,eV.\cite{Chen2006b, Consonni2009, Lefebvre2011, Nogues2014, Laehnemann2014}

The inset of Fig.~\ref{fig:fig2} depicts the TRPL transient of the ($D^0,X_{\text{A}}$) transition integrated over a 3\,meV wide spectral window centered on the transition (indicated by the dashed lines in the spectrum). An effective decay time of $\tau_{\text{eff}}= (0.16\pm 0.02)$\,ns is extracted from a fit of the data with a single exponential decay within the first ns convoluted with the system response function (solid line). After about 1\,ns, the $X_{\text{A}}$ and ($D^0,X_{\text{A}}$) transients deviate from a pure exponential behavior due to the coupling between these states and deeper 
states such as acceptor-bound excitons or the ($U,X$) states.\cite{Hauswald2013} The time of 0.16\,ns accounting for the initial two orders of magnitude of the decay is much shorter than expected for a purely radiative recombination of the ($D^0,X_{\text{A}}$) complex in free-standing, bulk-like GaN, for which values in excess of 1\,ns have been measured.\cite{Monemar2008,Monemar2010} However, similar values for the ($D^0,X_{\text{A}}$) decay time in GaN NWs have been reported by different groups,\cite{Calleja2000, Schlager2008a, Corfdir2009d, Park2009, Park2010, Gorgis2012} while no agreement has been reached on whether this decay time signifies a radiative or nonradiative decay of this state. Prior to commenting on this point, let us stress that the key for the understanding of the exciton dynamics in GaN NWs turns out to be the fact that the $X_{\text{A}}$ and ($D^0,X_{\text{A}}$) transitions exhibit identical decay times as illustrated by the dashed line (guide to the eye) in the inset of Fig.~\ref{fig:fig2}. We have also confirmed the parallel decay by performing a spectral deconvolution of the $X_{\text{A}}$ and ($D^0,X_{\text{A}}$) transitions using the method introduced in our previous study.\cite{Hauswald2013} This common decay time suggests a strong coupling between these two exciton states as observed and discussed in  previous studies for GaN NWs,\cite{Corfdir2009d, Hauswald2013} and planar layers.\cite{Corfdir2009} Further evidence for the presence of this coupling will be provided in the following section.

\begin{figure*}[t]
   \includegraphics[width=0.66\textwidth]{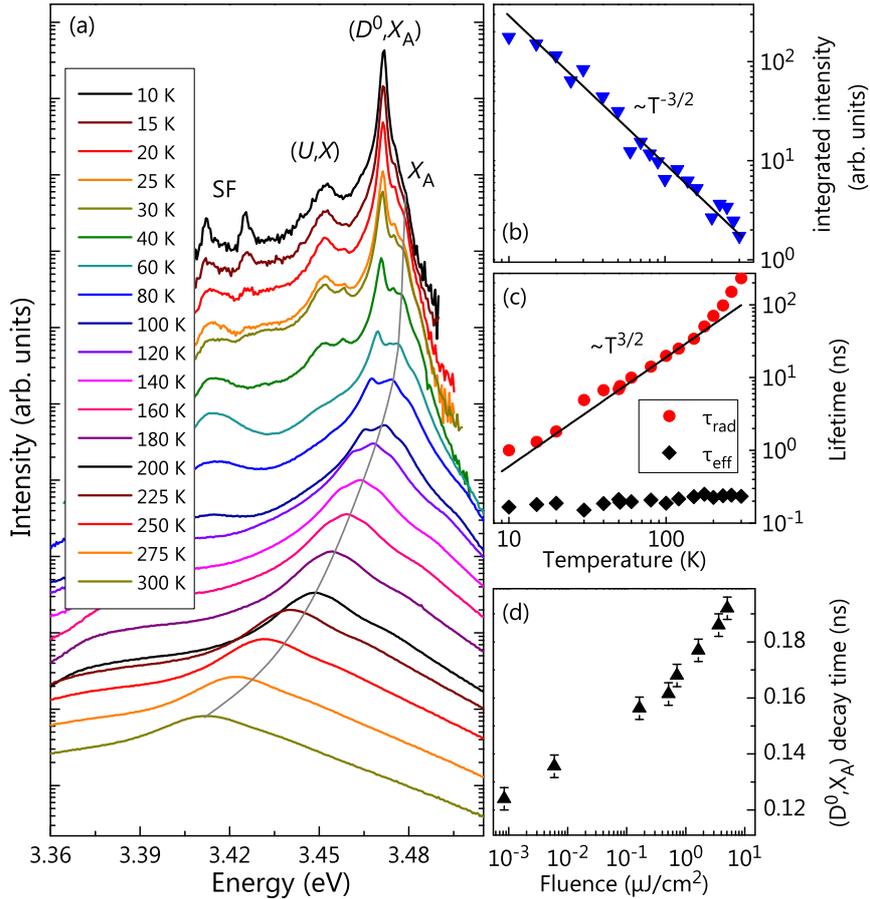}
   \caption{\label{fig:fig3}(color online) (a) Temperature-dependent cw PL spectra of sample R (the spectra have been shifted vertically for clarity). With increasing temperature, the bound exciton states are progressively delocalized, until the spectrum is dominated by the recombination of free A excitons ($X_A$) at room temperature (300\,K). The grey line is a guide to the eye and highlights the spectral shift of the X$_A$ transition with increasing temperature $T$. (b) Evolution of the spectrally integrated PL intensity of the donor-bound and free exciton states between 10 and 300\,K. The straight line shows that the integrated intensity decreases proportional to $T^{-3/2}$ with increasing $T$. (c) Effective ($\tau_{\text{eff}}$) and radiative ($\tau_{\text{rad}}$) lifetime obtained from temperature-dependent PL transients. The calculated values for $\tau_{\text{rad}}$ increase proportional to $T^{3/2}$, while $\tau_{\text{eff}}$ remains almost constant between 10 and 300\,K. (d) Dependence of $\tau_{\text{eff}}$ on the excitation fluence at $T=10$\,K. }
\end{figure*} 
Viewing a GaN NW with its flat top facet as a nano-cavity, one could argue that the spontaneous emission rate may be enhanced by the Purcell effect\cite{Purcell1946} assuming that the NW cavity supports a mode (i) matching the frequency of the investigated transition and (ii) having a spatial overlap with the position of the emitter.\cite{Maslov2006a} Quantitatively, however, a Purcell factor $f_P > 6$ would be required in order to enhance the radiative rate from 1\,ns$^{-1}$ in the bulk\cite{Monemar2008,Monemar2010} to 6\,ns$^{-1}$ in the NWs. Yet, calculated and experimentally determined values of $f_P$ averaged over the random position of the emitters in NWs are in the range of $1<f_P<2$.\cite{Kolper2011a, Paniagua-Dominguez2013a, Bleuse2011} In addition, for the Purcell effect to result in the mono-exponential decay over roughly two orders of magnitude observed experimentally (cf.\ inset of Fig.~\ref{fig:fig2}), where more than 100 NWs are probed simultaneously, $f_P$ would have to be essentially identical for the vast majority of NWs despite their highly irregular shapes [cf.\ Fig.~\ref{fig:fig1}(a)] and the variations in their length ($\pm 100$\,nm). It thus seems exceedingly unlikely that the short decay time of the ($D^0,X_{\text{A}}$) complex is related to a purely radiative process. The temperature-dependent PL and TRPL measurements presented in the following section confirm this conclusion and demonstrate unequivocally that the exciton decay in the sample under investigation is caused by a nonradiative process.

\subsection{Temperature dependence of the PL intensity and exciton lifetimes}
Figure \ref{fig:fig3}(a) displays the evolution of the cw PL spectra of sample R with increasing cryostat temperature $T$. The vertically shifted PL spectra show that the bound excitons are progressively delocalized according to their binding energy. Hence the relative contribution from the recombination of free A excitons is continuously increasing. At 300\,K, the spectrum is eventually dominated by the recombination of free A excitons at 3.41\,eV. In Fig.~\ref{fig:fig3}(b), the cw PL intensity of the bound and free exciton states determined by a lineshape fit is plotted versus $T$. The PL intensity starts to quench already upon an increase of $T$ to 15\,K (evidencing the presence of a nonradiative channel even for the lowest temperature) and follows a $T^{-3/2}$ dependence between 15 and 300\,K.  

In order to understand the origin of this quenching, we have performed temperature-dependent TRPL measurements. Figure~\ref{fig:fig3}(c) depicts the effective decay time $\tau_{\text{eff}}$ (black squares) of the ($D^0,X_{\text{A}}$) and $X_{\text{A}}$ transitions extracted from single-exponential fits to the short component of the PL transient integrated spectrally over both transitions. This procedure is justified since the two states exhibit a common lifetime already at 10\,K (cf.\ inset of Fig.~\ref{fig:fig2}). We find that $\tau_{\text{eff}}$ does not decrease with increasing temperature, but instead slightly increases from 0.16\,ns at $T=10\,$K to 0.23\,ns at $T=300\,$K. Figure~\ref{fig:fig3}(c) also shows the evolution of the radiative lifetime $\tau_{\text{rad}}$ (red circles) with increasing temperature normalized to 1\,ns at 10\,K.\cite{Monemar2008,Monemar2010} As the excitation pulse is much shorter than the recombination time, $\tau_{\text{rad}}$ is proportional to the inverse of the spectrally integrated PL peak intensity [cf. Eq.~(A6) in Ref.~\onlinecite{Brandt2002a}]. The radiative lifetime is proportional to $T^{3/2}$ as expected for free excitons from theoretical grounds.\cite{Andreani1992} The quenching of the PL intensity [cf.\ in Fig.~\ref{fig:fig3}(b)] is thus caused predominantly by the increase in the radiative lifetime, since the effective lifetime is essentially constant and equal to the nonradiative lifetime [cf.\ Fig.~\ref{fig:fig3}(c)]. Note that the increase of the radiative lifetime for temperatures above $10$\,K supports our conclusion from Sec.~\ref{sec:LTPL} that the $X_{\text{A}}$ and ($D^0,X_{\text{A}}$) states are already coupled in this regime. The deviation from the $T^{3/2}$ dependence at temperatures above 200\,K is due to the increasing participation of free carriers in recombination.\cite{Brandt1998b} 

Finally, Fig.~\ref{fig:fig3}(d) shows the dependence of the effective lifetime of the ($D^0,X_{\text{A}}$) on the fluence of the pulsed excitation at 10\,K. The effective lifetime $\tau_{\text{eff}}$ is found to increase by less than a factor of two when the excitation fluence is varied over roughly four orders of magnitude. The data presented in Fig.~\ref{fig:fig3}(d) suggest that the nonradiative channel responsible for the quenching of the PL intensity and the short effective lifetimes is highly efficient and cannot be saturated under the experimental conditions employed. The weak dependence guarantees that different excitation densities caused, for example, by a variation of the area fill factor for different NW ensembles, do not lead to pronounced changes in $\tau_{\text{eff}}$. This result therefore allows us to compare the effective lifetime of GaN NW ensembles with different morphological properties and will be used in Sec.~\ref{sec:overview} to systematically investigate the dependence of $\tau_{\text{eff}}$ on the surface-to-volume ratio and the coalescence degree.  

The combination of the experimental results presented in Fig.~\ref{fig:fig3} shows unambiguously that the effective lifetime of the coupled system ($D^0,X_{\text{A}}$) $\leftrightarrows$ $X_{\text{A}} + D^0$ is dominated by a nonradiative decay channel over the whole investigated temperature range. Furthermore, the results imply that the free and bound exciton states are coupled at lower temperatures than expected from the binding energy of the ($D^0,X_{\text{A}}$) complex. This finding corroborates the results from our previous work, where the inevitable presence of electric fields in GaN NWs was suggested as the cause of an enhanced coupling between all excitonic states.\cite{Hauswald2013} In order to gain a deeper understanding of the consequences of this coupling and to aid the interpretation of the effective decay rate of the system as a whole, we proceed by investigating a rate equation system describing the dynamics of the two coupled states.

\subsection{Rate equation model describing the coupling at low temperatures}
\begin{figure}[t]
   \includegraphics*[width=8.5cm]{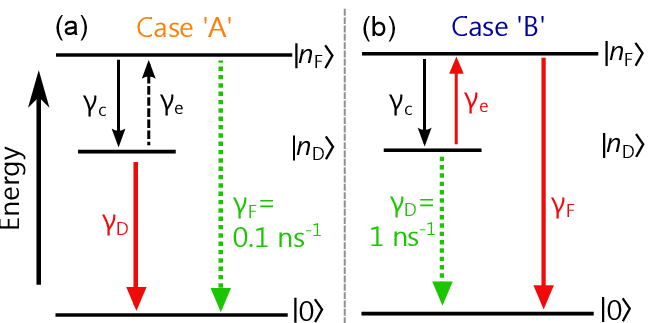}
   \includegraphics*[width=8.5cm]{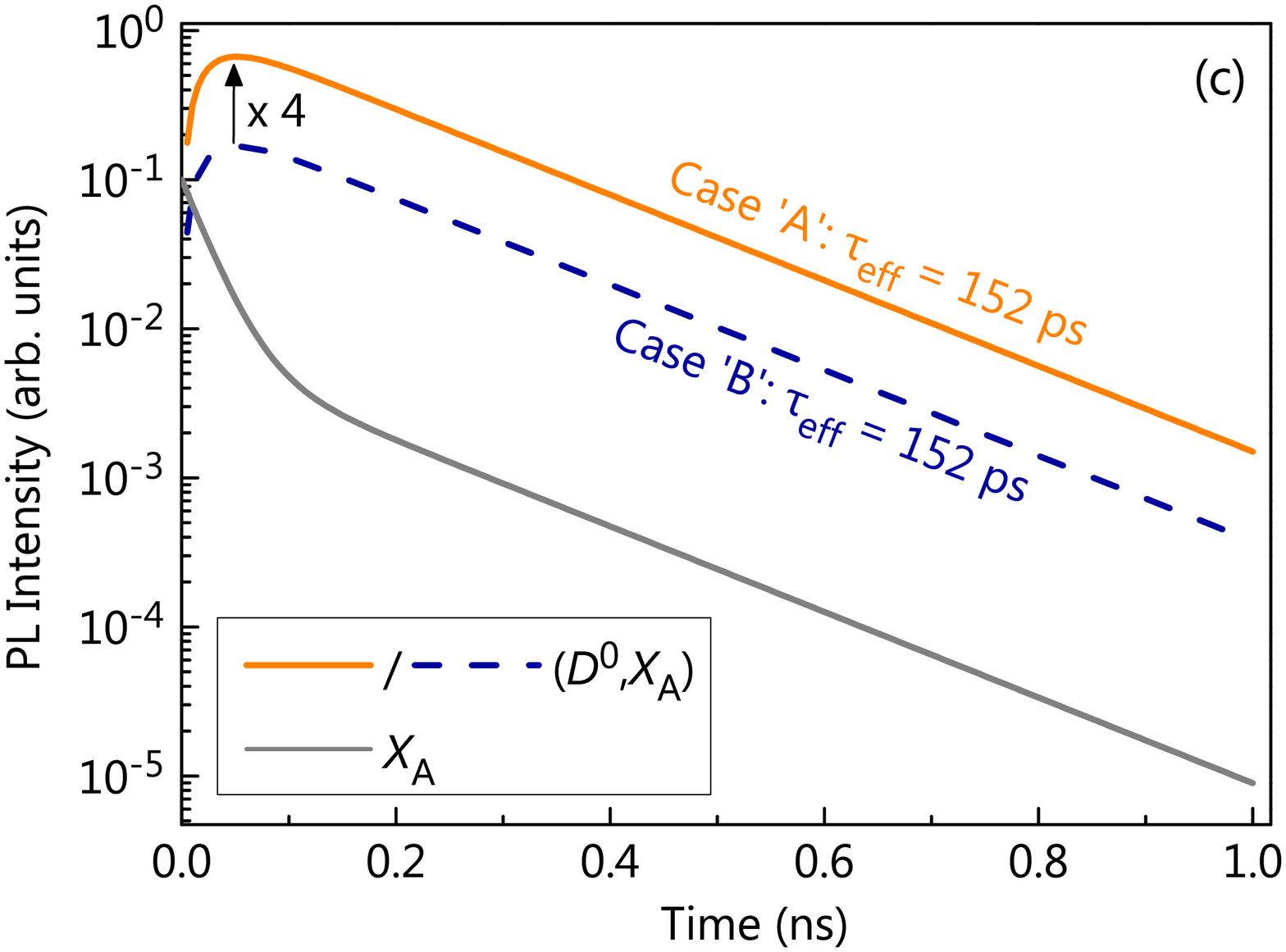}
   \caption{\label{fig:fig4}(color online) (a) and (b) Schematic energy diagrams visualizing Eqs.~(\ref{model1}) and (\ref{model2}) for the two different cases 'A' and 'B', respectively. The two states involved are denoted by $| n_{\text{i}} \rangle$, and the crystal ground state is represented by $| 0 \rangle$. The red solid arrows symbolize the effective decay channel, which dictates the dynamics of the coupled system, while the green dotted arrows represent purely radiative transitions. (c) Simulated TRPL transients of the ($D^0,X_{\text{A}}$) and $X_{\text{A}}$ given by $\gamma_{\hspace{0.3mm}\text{i,r}}n_{\text{i}}$, respectively, employing the effective decay rates $\gamma_{\hspace{0.3mm}\text{i}}$ as summarized in Table \ref{tab:parameters} and radiative decay rates $\gamma_{\hspace{0.3mm}\text{i,r}}$ for the free and bound exciton of 0.1 and 1\,ns$^{-1}$, respectively.} 
   \end{figure}

The coupled ($D^0,X_{\text{A}}$) $\leftrightarrows$ $X_{\text{A}} + D^0$ system, for which an energy scheme is displayed in Figs.~\ref{fig:fig4}(a) and \ref{fig:fig4}(b), is described by the following set of linear rate equations in the limit of low excitation density:
\begin{align}
\label{model1}
\frac{dn_{\text{F}}}{dt} & = -\gamma_{\hspace{0.3mm}\text{c}} n_{\text{F}} + \gamma_{\hspace{0.3mm}\text{e}} n_{\text{D}} - \gamma_{\hspace{0.3mm}\text{F}} n_{\text{F}} \\
\label{model2}
\frac{dn_{\text{D}}}{dt} & = \gamma_{\hspace{0.3mm}\text{c}} n_{\text{F}} - \gamma_{\hspace{0.3mm}\text{e}} n_{\text{D}} - \gamma_{\hspace{0.3mm}\text{D}} n_{\text{D}}. 
\end{align}
Here, $n_{\text{F}}$ and $n_{\text{D}}$ denote the time-dependent densities of free and donor-bound excitons with the initial densities $n_{\text{F}}(0)=n_{\text{F}}^0$ and  $n_{\text{D}}(0)=0$, respectively, while $\gamma_{\hspace{0.3mm}\text{F}}$ and $\gamma_{\hspace{0.3mm}\text{D}}$ represent the sums of the respective radiative ($\gamma_{\hspace{0.3mm}\text{i,r}}$) and nonradiative rates ($\gamma_{\hspace{0.3mm}\text{i,nr}}$) with i\,$=\{\text{F,\,D}\}$. The coupling is represented by $\gamma_{\hspace{0.3mm}\text{c}}$, the capture rate of free to bound excitons, and $\gamma_{\hspace{0.3mm}\text{e}}$, the rate for the dissociation of the ($D^0,X_{\text{A}}$) complex and the subsequent return of the exciton to its free state.

This rate equation system can be solved analytically, which yields the time-dependent densities for both states:

\begin{align}
\label{sol1}
n_{\text{F}}(t) &= \frac{n_{\text{F}}^0}{2\beta} \left[\left(\beta-\delta\right)\mathrm e^{-\frac{1}{2}(\alpha-\beta)t}+\left(\beta+\delta\right)\mathrm e^{-\frac{1}{2}(\alpha+\beta)t}\right]\\
\label{sol2}
n_{\text{D}}(t) &= \frac{\gamma_{\hspace{0.3mm}\text{c}} n_{\text{F}}^0}{\beta} \left[\mathrm e^{-\frac{1}{2}(\alpha-\beta)t}-\mathrm e^{-\frac{1}{2}(\alpha+\beta)t}\right],
\end{align}

with

\begin{align}
\label{Eq:alpha}
\alpha &= \gamma_{\hspace{0.3mm}\text{c}}+\gamma_{\hspace{0.3mm}\text{e}}+\gamma_{\hspace{0.3mm}\text{D}}+\gamma_{\hspace{0.3mm}\text{F}},\\
\label{Eq:beta}
\beta &= \sqrt{\alpha^2-4\left[\gamma_{\hspace{0.3mm}\text{e}}\gamma_{\hspace{0.3mm}\text{F}}+\gamma_{\hspace{0.3mm}\text{D}}\left(\gamma_{\hspace{0.3mm}\text{c}}+\gamma_{\hspace{0.3mm}\text{F}}\right)\right]},\\
\label{Eq:delta}
\delta &= \gamma_{\hspace{0.3mm}\text{c}}-\gamma_{\hspace{0.3mm}\text{e}}-\gamma_{\hspace{0.3mm}\text{D}}+\gamma_{\hspace{0.3mm}\text{F}}.
\end{align}

The time-dependent densities [Eqs.~(\ref{sol1})--(\ref{sol2})] are governed by two exponentials with different decay constants. The terms proportional to exp$\left[-\frac{1}{2}(\alpha+\beta)t\right]$ describe the increase of the donor-bound exciton population $n_{\text{D}}$ and the simultaneous fast decrease of the free exciton population $n_{\text{F}}$ directly after excitation. At longer times, the system reveals its coupled nature by exhibiting a common effective decay rate, which is represented by the terms proportional to exp$\left[-\frac{1}{2}(\alpha-\beta)t\right]$ in Eqs.~(\ref{sol1}) and (\ref{sol2}).

If donor-bound and free excitons do not couple (i.\,e., $\gamma_{\hspace{0.3mm}\text{e}}=0$), the common decay rate of both states naturally disappears since $\left|\beta\right|- \left|\delta\right| =0$, and the effective decay rates of the free and bound exciton states are given by $\gamma_{\hspace{0.3mm}\text{c}}$+$\gamma_{\hspace{0.3mm}\text{F}}$ and $\gamma_{\hspace{0.3mm}\text{D}}$, respectively, as expected for the uncoupled system.

The dynamics of the coupled system ($D^0,X_{\text{A}}$) $\leftrightarrows$ $X_{\text{A}} + D^0$ is invariant with respect to a transformation $(\gamma_{\hspace{0.3mm}\text{c}},\gamma_{\hspace{0.3mm}\text{e}},\gamma_{\hspace{0.3mm}\text{D}},\gamma_{\hspace{0.3mm}\text{F}})\rightarrow (\gamma\,_\text{c}^{\prime},\gamma\,_\text{e}^{\prime},\gamma\,_\text{D}^{\prime},\gamma\,_\text{F}^{\prime})$, if $\alpha$, $\beta$, and $\delta$ remain constant. Such a transformation always exists, and we will now discuss a particular example representing two limiting cases.

Let us first consider the situation depicted in Fig.~\ref{fig:fig4}(a) (hereafter referred to as 'case A'). The free exciton is assumed to have an effective decay rate of $\gamma_{\hspace{0.3mm}\text{F}} = \gamma_{\hspace{0.3mm}\text{F,r}} = 0.1$\,ns$^{-1}$, i.\,e., equal to its radiative decay rate $\gamma_{\hspace{0.3mm}\text{F,r}}$  estimated in Ref.~\onlinecite{Korona2002}. To account for the experimentally observed rapid decay, the effective decay rate of the donor-bound exciton is set to 7\,ns$^{-1}$, i.e.\,, its decay is mostly nonradiative. The rates determining the coupling are set to $\gamma_{\hspace{0.3mm}\text{c}} = 40$\,ns$^{-1}$ and $\gamma_{\hspace{0.3mm}\text{e}} = 2$\,ns$^{-1}$ representing a comparatively weak coupling (i.e.\ $\gamma_{\hspace{0.3mm}\text{e}} \ll \gamma_{\hspace{0.3mm}\text{c}}$) between the two states as derived from the analysis of TRPL data in our previous work.\cite{Hauswald2013} The resulting time-dependent intensity $\gamma_{\hspace{0.3mm}\text{F,r}}n_{\text{F}}(t)$ of the free exciton and $\gamma_{\hspace{0.3mm}\text{D,r}}n_{\text{D}}(t)$ of the donor-bound exciton are depicted in Fig.~\ref{fig:fig4}(c). Already 0.1\,ns after the generation of free excitons, the ($D^0,X_{\text{A}}$) and $X_{\text{A}}$ states decay with a common lifetime of about 0.15\,ns, in agreement with the experimental result presented earlier [cf.\ inset of Fig.~\ref{fig:fig2}]. Note that the time resolution of 50\,ps impedes the observation of the fast capture process of the free excitons.

\begin{table}[b]
\caption{Parameters used in the rate equation model [Eqs.~(\ref{model1})--(\ref{model2})] to compute the PL transients shown in Fig.~\ref{fig:fig4}(c), and the effective rate constants given by Eqs.~(\ref{sol1})--(\ref{sol2}), all in ns$^{-1}$.}
\begin{ruledtabular}
\begin{tabular}{lccccccc}
 & $\gamma_{\hspace{0.3mm}\text{F}}$ & $\gamma_{\hspace{0.3mm}\text{D}}$ & $\gamma_{\hspace{0.3mm}\text{c}}$ & $\gamma_{\hspace{0.3mm}\text{e}}$ & $\alpha$ & $\beta$ & $\delta$ \\
\colrule 
Case 'A'  & 0.1 & 7  & 40 & 2 & 49.1 & 35.9 & 31.1\\
Case 'B'  & 30.1 & 1  & 10 & 8 & 49.1 & 35.9 & 31.1\\
\end{tabular}
\end{ruledtabular}
\label{tab:parameters}
\end{table}

Next, we assume that the ($D^0,X_{\text{A}}$) decay is purely radiative, implying that $\gamma\,_\text{D}^{\prime} = \gamma_{\hspace{0.3mm}\text{D,r}} = 1$\,ns$^{-1}$ (hereafter referred to as 'case B').\cite{Monemar2008,Monemar2010} We also demand that the effective rate constants $\alpha$, $\beta$, and $\delta$ remain constant, which yields three algebraic equations for the three remaining rates $\gamma\,_\text{c}^{\prime}$, $\gamma\,_\text{e}^{\prime}$, and $\gamma\,_\text{F}^{\prime}$. The resulting equation system has one and only one solution, namely, that given in Table~\ref{tab:parameters}. As expected from the equal values for the effective rate constants $\alpha$, $\beta$, and $\delta$, the decay times obtained are identical for cases 'A' and 'B' as illustrated by the simulated transients in Fig.~\ref{fig:fig4}(c). However, the system of case 'B' is now strongly coupled (i.e.\ $\gamma_{\hspace{0.3mm}\text{e}} \lesssim \gamma_{\hspace{0.3mm}\text{c}}$), and \emph{the nonradiative recombination takes place indirectly via the free exciton state} as $\gamma_{\hspace{0.3mm}\text{F}}\approx \gamma_{\hspace{0.3mm}\text{F,nr}}=30.1\,$ns$^{-1}$ [cf. Fig.~\ref{fig:fig4}(b)].    

The two limiting cases 'A' and 'B' are mathematically distinguishable, since the intensity ratio of the free and bound exciton transitions changes. This change is due to the fact that $n_D(t)$ is proportional to $\gamma_{\hspace{0.3mm}\text{c}}$. However, distinguishing cases 'A' and 'B' (or any of the infinitely many solutions between these two limiting extremes of purely radiative recombination of either the free or the bound exciton) will be difficult in practice, if not impossible. The reason for this fact is that the experimentally accessible PL intensity is proportional to the \emph{external} quantum efficiency. The intensity ratio of the free and bound exciton transitions is thus influenced by the different extraction efficiencies for the emission from free and bound excitons, which in turn are caused largely by the reabsorption of the free-exciton emission, but also by the different spatial distributions of the emitters.\cite{Maslov2006a} Additionally, the collection efficiency will be affected by the different far-field patterns of the respective modes within the NWs.\cite{Maslov2004, Motohisa2014} 

In general, we will thus not be able to determine the individual decay rates of the states participating in recombination, but are left with the effective decay rate of the coupled system. In particular, the coupled system may decay via either of its constituent states (or even a combination of both) without giving the experimentalist the possibility to decide which path was actually taken. In other words, the ($D^0,X_{\text{A}}$) $\leftrightarrows$ $X_{\text{A}} + D^0$ coupled system should be viewed as one entity, and a distinction between the free and bound states based on their dynamics is not physically meaningful. 
\begin{figure*}[t]
   \includegraphics[width=0.99\textwidth]{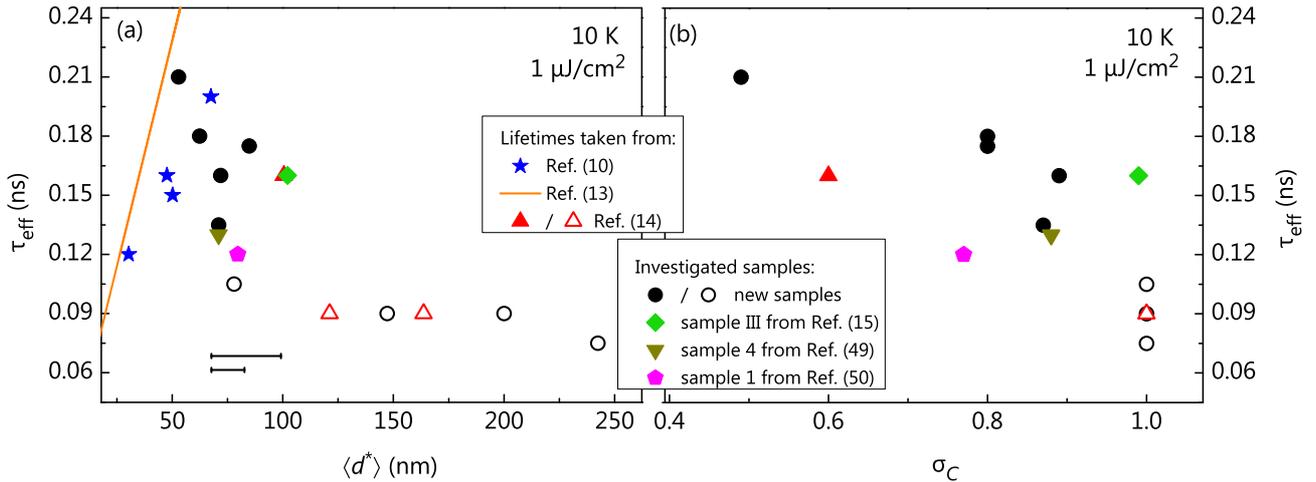}
  \caption{\label{fig:fig5}(color online) Overview of the effective ($D^0,X_{\text{A}}$) lifetime obtained by fitting the short component of the respective TRPL transient recorded at low temperatures with a single-exponential decay for 19 different GaN NW ensembles. Solid symbols represent samples grown using the self-induced growth mode, while the open symbols are samples grown by SAG.\cite{Schumann2011b} (a) Dependence of the effective lifetime $\tau_{\text{eff}}$ on the parameter $\langle d^*\rangle$. The solid line depicts the expected lifetimes for an effective surface recombination velocity of $\textit{\~{S}}=6.3\times 10^3$\,cm/s.\cite{Gorgis2012} The horizontal bars indicate the typical standard deviation of the $d^*$-distribution in the NW ensemble for the self-induced growth (upper bar) and SAG (lower bar). (b) Dependence of $\tau_{\text{eff}}$ on the coalescence degree $\sigma_C$ of the sample.}
\end{figure*} 

\subsection{Dependence of the effective lifetime on surface-to-volume ratio and coalescence degree}
\label{sec:overview}

The above discussion has important consequences, particularly with regard to the question about the origin of the nonradiative channel dominating the exciton decay in GaN NWs. In case 'A', the wave function of the majority of donor-bound excitons has to overlap with that of a nonradiative center to explain the experimentally observed transients, thus requiring fairly high densities of these centers since the ($D^0,X_{\text{A}}$) complex is a localized state with a small spatial extension.\cite{Hauswald2014} In case 'B', in contrast, the nonradiative decay occurs via the free exciton, which is principally a coherent excitation of the entire NW,\cite{Pfueller2010b} but is also from a classical point of view  a spatially delocalized particle, which experiences its surroundings on the scale of its diffusion length of 50--200\,nm.\cite{Speck1999, Ino2008a, Nogues2014} Considering that these values are actually larger than the radii of the NWs under investigation, it is interesting to compile the decay times of the ($D^0,X_{\text{A}}$) state for different GaN NW samples and examine the data for a correlation with the mean surface-to-volume ratio and the coalescence degree of the ensemble.  

Assuming a finite surface recombination velocity $S$, the surface recombination rate in a single NW with equivalent disk diameter $d_\text{disk}$ is often approximated by the relation 
\begin{align}
\gamma_{\hspace{0.3mm}\text{S}}=\frac{4S}{d_\text{disk}},
\end{align}
which is strictly valid for cylindrical objects only. However, the inadvertent coalescence of individual NWs during growth distorts their shape from a regular hexagon towards elongated and branched shapes [cf. Fig.~\ref{fig:fig1}(a)]. Hence, the equivalent-disk diameter $d_{\text{disk}}$ systematically underestimates the influence of the surface for NWs exhibiting a low circularity. For an arbitrarily shaped object, we show in Appendix \ref{sec:appendixA} that the surface recombination rate $\gamma_{\hspace{0.3mm}\text{S}}$ is actually proportional to the perimeter-over-area ratio:

\begin{align}
\label{Eq:area-perimeter}
\gamma_{\hspace{0.3mm}\text{S}}=\frac{SP}{A}=\frac{4S}{d^*},
\end{align}
where $d^*=4A/P$, with the cross-sectional area $A$ of a NW and the perimeter $P$ of its cross section. The parameter $d^*\leq d_{\text{disk}}$ thus represents a new effective diameter for a given NW taking into account the larger perimeter of coalesced NWs compared to uncoalesced ones when both exhibit the same cross-sectional area $A$. Since we investigate NW ensembles rather than single NWs in this study, we employ the mean of the respective distribution to characterize the ensemble. For sample R, the distribution of $d^*$ (not shown) is symmetric and is described well by a normal distribution with a mean of $\langle d^*\rangle=72$\,nm. This value is 20\% smaller than the mean equivalent disk diameter $\langle d_{\text{disk}}\rangle=99$\,nm [cf.\ Fig.~\ref{fig:fig1}(b)], underlining the importance of considering the actual shape of the NWs for the quantitative analysis of surface-related phenomena.

We are now in the position to quantitatively compare the effective lifetime of the coupled ($D^0,X_{\text{A}})\leftrightarrows X_{\text{A}}$ system for a number of GaN NW samples with different values of $\langle d^*\rangle$ grown in several different MBE machines using different plasma sources. The results based on samples from Refs.~\onlinecite{Hauswald2014}, \onlinecite{Dogan2011a}, and \onlinecite{Jenichen2011a} are displayed in Fig.~\ref{fig:fig5}(a) together with data reported in Refs.~\onlinecite{Corfdir2009d} and \onlinecite{Hauswald2013}. The effective lifetimes measured for these samples remain short compared with the radiative lifetime regardless of the value of $\langle d^* \rangle$. If the effective lifetime $\tau_{\text{eff}}$ were fundamentally limited by surface recombination, the resulting values would be expected to linearly increase with $\langle d^* \rangle$. The solid line shows the expected dependence with $\textit{\~{S}} = 6.32\times 10^{3}\,$cm/s as reported by \citet{Gorgis2012} for bound excitons. Note that we have not yet observed decay times longer than expected from this value, but only shorter ones. For thin NWs, surface recombination may therefore be the limiting process after all, but for thicker ones, it is clear that a process not related to the surface determines the exciton decay rate. Figure~\ref{fig:fig5}(b) plots the effective lifetime $\tau_{\text{eff}}$ from 15 different GaN NW samples versus their coalescence degree $\sigma_C$. With the exception of the GaN NW arrays fabricated by SAG (whose coalescence degree was set to $\sigma_C=1$ since multiple NW nucleations occur inevitably for each hole~\cite{Schumann2011b}), the values of the effective exciton lifetime scatter around 150\,ps and do not show any obvious trend towards shorter values for higher coalescence degrees. 

\section{Conclusion}
Our results show conclusively that the exciton lifetime in the investigated GaN NW ensembles is limited by nonradiative recombination. The lifetime is independent of the ensembles mean surface-to-volume ratio and coalescence degree, implying that the nonradiative process is neither caused by surface recombination nor by dislocations formed due to NW coalescence. The remaining possibility for the origin of this nonradiative channel are point defects. In fact, given that the substrate temperatures used in the MBE growth of GaN NWs are low compared to those used in both metal-organic and hydride vapor phase epitaxy, the point defect density in these NWs is necessarily higher than in state-of-the-art GaN layers, which are grown closer to equilibrium. However, the point defect density does not need to be excessively high in order to dominate the exciton decay. We have demonstrated that bound and free excitons in the investigated GaN NWs are strongly coupled even at low temperatures, and the nonradiative decay may thus take place via the free state. Given that the diffusion length of free excitons in GaN is larger than 50\,nm, a density as low as 10$^{15}$\,cm$^{-3}$ may suffice to introduce an effective nonradiative decay channel. 

Attempts to reduce the density of this detrimental defects would probably require a higher substrate temperature. However, due to an exponential increase of both the desorption of Ga adatoms and the dissociation of GaN in the high vacuum environment of MBE, it seems unlikely that sufficiently high substrate temperatures can be reached. A perhaps more promising approach for reducing the impact of these defects at least at low temperatures is the manipulation of the strength of the coupling between free and bound states in the NWs. As discussed in detail in our previous work,\cite{Hauswald2013} we believe that the enhanced coupling in GaN NWs is nonthermal in nature and related to the presence of electric fields within the NWs. Since these electric fields arise from the pinning of the Fermi level at the NW sidewalls, a passivation of the corresponding \emph{M}-plane surfaces may drastically diminish the coupling and result in a higher internal quantum efficiency at low temperatures.   

\section*{Acknowledgment}
The authors would like to thank P.~Dogan and M.~Knelangen for providing additional samples for our study, A.-K. Bluhm for the scanning electron micrographs used in Fig.~\ref{fig:fig5}, L.~Schrottke for a critical reading of the manuscript, and H.~Riechert for continuous support. We gratefully acknowledge partial financial support of this work by the Deutsche Forschungsgemeinschaft through SFB 951. K.~K.~Sabelfeld kindly acknowledges the support of the Russian Science Foundation under RScF Grant 14-11-00083.

\appendix
\section{Surface recombination for arbitrary NW shapes}

\label{sec:appendixA}

Let us consider generation, diffusion, and recombination of excitons described by the equation
\begin{equation}
\frac{\partial n(\mathbf{x},t)}{\partial t}=D\Delta n(\mathbf{x},t)-\gamma n(\mathbf{x},t)+G\delta(t)\label{eq:1}
\end{equation}
within a domain $\mathcal{A}$, with the initial condition $n(\mathbf{x},0)=0$ and the boundary condition at the border of the domain 
\begin{equation}
D\frac{\partial n}{\partial\nu}(\xi)+Sn(\xi)=0,\,\,\,\xi\in\partial\!\mathcal{A}.\label{eq:2}
\end{equation}
Here $n(\mathbf{x},t)$ is the exciton density, $G$ the generation rate, $D$ the diffusion coefficient, $\gamma$ the radiative recombination rate, $S$ the surface recombination velocity, and $\nu$ is a unit normal directed outwards of the domain.

Our aim is to calculate the PL intensity (normalized to the generation
rate $G$ and the domain area $A$)
\begin{equation}
I(t)\equiv\frac{1}{GA}\int_\mathcal{A} n(\mathbf{x},t)d\mathbf{x}\label{eq:3}
\end{equation}
in the limit of large diffusivity, i.e.\ , we thus require that the diffusion length $\sqrt{D/\gamma}$ is large compared to the domain size so that $D/\gamma\gg A$. The analysis does not depend on the dimensionality of the system, and we refer below to the domain area $A$ and its perimeter $P$ in application to the two-dimensional
problem considered in this manuscript. The results presented below are applicable
to a $d$-dimensional volume $A$ and its $(d-1)$-dimensional hypersurface
$P$.

Since the diffusion coefficient $D$ is large, we put $D=D_{0}/\varepsilon$,
where $\varepsilon$ is a small quantity. Then, the diffusion equation
(\ref{eq:1}) and the boundary condition (\ref{eq:2}) read
\begin{equation}
\varepsilon\frac{\partial n(\mathbf{x},t)}{\partial t}=D_{0}\Delta n(\mathbf{x},t)-\varepsilon\gamma n(\mathbf{x},t)+\varepsilon G\delta(t),\label{eq:4}
\end{equation}
\begin{equation}
D_{0}\frac{\partial n}{\partial\nu}(\xi)+\varepsilon Sn(\xi)=0,\,\,\,\xi\in\partial\! \mathcal{A}.\label{eq:5}
\end{equation}
Let us expand $n(\mathbf{x},t)$ in a power series over $\varepsilon$, $n(\mathbf{x},t)=n_{0}+\varepsilon n_{1}+\ldots.$ Substituting this expansion in Eqs.~(\ref{eq:4}) and (\ref{eq:5}), we obtain the first two terms over the powers of $\varepsilon$ as 
\begin{equation}
D_{0}\Delta n_{0}(\mathbf{x},t)=0,\,\,\mathbf{x}\in \mathcal{A},\,\,\,\,\,\,\, D_{0}\frac{\partial n_{0}}{\partial\nu}(\xi)=0,\,\,\xi\in\partial\!\mathcal{A},\label{eq:6}
\end{equation}
and
\begin{equation}
\frac{\partial n_{0}(\mathbf{x},t)}{\partial t}=D_{0}\Delta n_{1}(\mathbf{x},t)-\gamma n_{0}(\mathbf{x},t)+G\delta(t),\,\,\mathbf{x}\in \mathcal{A},\label{eq:7}
\end{equation}
\begin{equation}
D_{0}\frac{\partial n_{1}}{\partial\nu}(\xi)+Sn_{0}(\xi)=0,\,\,\xi\in\partial\!\mathcal{A}.\label{eq:8}
\end{equation}

It follows from Eq.~(\ref{eq:6}) that $n_{0}$ does not depend on $\mathbf{x}$ so that $n_{0}=n_{0}(t)$. Now we integrate Eq.~(\ref{eq:7}) over the domain area $A$:
\begin{equation}
A\frac{\partial n_{0}(t)}{\partial t}=D_{0}\int_{A}\Delta n_{1}d\mathbf{x}-A\gamma n_{0}(t)+AG\delta(t).\label{eq:9}
\end{equation}
Since
\[
\int_{\mathcal{A}}\Delta n_{1}d\mathbf{x}=\int_{\partial\!\mathcal{A}}\frac{\partial n_{1}}{\partial\nu}d\xi,
\]
we arrive using the boundary condition (\ref{eq:8}) at the equation 
\begin{equation}
\frac{\partial n_{0}(t)}{\partial t}=-\left(\gamma+\frac{SP}{A}\right)n_{0}(t)+G\delta(t).\label{eq:10}
\end{equation}
In the limit of an infinitely large diffusion coefficient, $n_{0}(t)$ no longer explicitly depends on $D$ and Eq.~(\ref{eq:10}) gives the desired result
\begin{equation}
I(t)=\exp\left[-\left(\gamma+\gamma_{\hspace{0.3mm}\text{S}}\right)t\right],\label{eq:11}
\end{equation}
where
\begin{equation}
\gamma_{\hspace{0.3mm}\text{S}}=\frac{SP}{A},\label{eq:12}
\end{equation}
is the surface recombination rate. An account of the subsequent terms in the expansion over $\varepsilon$ can give the PL intensity in the case of a large but finite diffusivity.

\end{document}